# Multifocal Terahertz Lens with Adjustable Focal Points

Ali Abdolali [1,*], Roya Kargar [1], and Kasra Rouhi [1]

[1] Applied Electromagnetic Laboratory, School of Electrical Engineering, Iran University of Science and Technology, Tehran, 1684613114, Iran

E-mail: abdolali@iust.ac.ir

**Abstract**

The conventional lens's tunability drawback always restricts their application compared to the metasurface lens (metalens). On the other side, reconfigurable metalenses offer the benefits of ultrathin thickness and capable of tunability. Therefore achieving reconfigurable functionalities in a single metasurface has attracted significant research interest for potential terahertz (THz) applications. In this paper, an adjustable metasurface is presented using Vanadium dioxide (VO2) to manipulate the electromagnetic waves and provide the full reflection phase. The phase-change metasurface is composed of a VO2 nanofilm, a silicon spacer, and a gold layer embedded in the structure's bottom. By employing the reconfigurable metasurface with the specific phase distribution, the incident beam can converge to determined points in any arbitrary manner, including the number of the focal points, focal points location, and power intensity ratio. Numerical simulations demonstrate that the proposed reconfigurable metasurface can concentrate power on one or more than one focal point in reflection modes as expected. Additionally, the VO2-based metasurface can control concentration width in a real-time manner using a novel proposed method. The simulation and theoretical results are in good agreement to verify the validity and feasibility of 2-bit metalens design, which has considerable potential in wireless high-speed communication and super-resolution imaging.

## 1. Introduction

Metasurface, referring to the thin planar arrangement of artificial unit cells in the subwavelength scale, has presented versatile abilities in manipulating the amplitude, phase, and polarization of electromagnetic waves for various applications [1]. A metasurface can realize the manipulation of transmission and reflection response by adjusting the planar unit cell configuration. Frequently, metasurface in optical and THz region offers a new paradigm for applications, including vortex beam [2], optical signal processors [3]–[5], absorber [6]–[8], wavefront manipulation [9], wave diffusion [10], and lens [11], [12]. Compared to conventional three-dimension and bulky wave manipulation methods, metasurfaces can be advantageously fabricated since their reduced profiles are more straightforward to realize by modern and simple fabrication techniques. With the growing demand for miniaturized and highly integrated systems, a single metasurface showing multiple functionalities is desirable for reconfigurable communication and imaging systems.

Among all these applications mentioned above, beam concentration, along with ultrathin characteristics, is an excellent request for its wide-spread application in microscopy, imaging, and spectroscopy [13]. As an indispensable tool, the optical lens, based on bulk optical components, is widely used in various electromagnetic devices. There is no doubt that a metasurface can miniaturize lenses to overcome the lack of big volume and frequency dispersion. Metalenses can give better optical functionalities and allow for a much more compact device design than the conventional high-end objective lenses [14]. Interestingly, several research groups also investigated the possibility of designing a lens that can control the focal point by adjusting the reflection/transmission phase of each unit cell individually [12], [15], [16]. In [17], the metasurface lens is designed to manipulate the radiated beam in the Cassegrain system. In [18], the authors introduce broadband unidirectional cloaks based on flat metasurface lenses to hide the cloak region. In other research [19], circularly polarization-sensitive metasurface has exhibited that it can control illuminated waves with specific polarization independently. Indeed it can control the transmitted phase for left-handed and right-handed circularly polarized waves and forward transmitted waves to the desired point. In [20], the authors provide a microlens array for imaging applications with high resolution. More importantly, in the mentioned researches, the lack of tunability dramatically decreases their practical applications. Due to increasing interest in system integration, a real-time concentrator with a tunable unit cell is highly desired, notably at THz frequencies [21], [22].

Procedures of including tunable materials into metasurfaces for tuning the functionality, such as the use of liquid crystals [23], phase-change material [24], Indium Tin Oxide [25], graphene [26]–[28], and pin-diode [29], [30], are wide-spread due to their degrees of freedom. These metasurface's dynamic response enables active THz metasurface, stimulated by external stimuli via photoexcitation, electric or magnetic bias, and temperature. In this work, in order to implement such a platform, we have benefited from VO2 unique properties that can be controlled by external stimulation. VO2 is a smart material that undergoes an ultrafast and brutal reversible first order insulator-to-metal transition, which can be controlled by some external parameters such as temperature [31], applied current [32], and optical pumping [33]. Atomic-level deformation in phase change material can offer drastic variation in material characteristics in a broad spectral range during the phase transition. This transition in VO2 can occur within an order of several nanoseconds or even in the picoseconds range for optical activities. These characteristics and almost critical temperature ($T_c = 68$ °C) make VO2 film suitable for many encouraging applications in many fields.

Recently, we proposed a one-bit tunable metalens based on coding metasurface that two reflective unit cells with 180º phase differences were employed to achieve digital metamaterials [21]. Nevertheless, in the previous paper, the one-bit metalens' performance is restricted because of a limited number of possible states for the reflection phase. Also, the intensity ratio of the focal points was not adjustable. To achieve multifocal spots in a different position, it undergoes a major problem where the two simultaneously focal points are placed at the same axis normal to the structure. This paper presents a design procedure for near-field coding metasurface with single and multi-focus characteristics. In order to provide the full reflection phase of 360º, the VO2 film is employed in the adjustable metasurface. The Huygens metasurface presence contributes a systematic theoretical foundation to perform the manipulation of reflection. Furthermore, multi-focus Huygen metalens is rendered to converged the incident beam to any arbitrary focal points at 2.05 THz. Besides, our numerical results agree very well with theoretical results, which adds a new degree of freedom to manipulate electromagnetic waves. On the other hand, full width at half maximum (FWHM) is also an important parameter commonly used to describe the focal point's width. The amount of this parameter can be varied according to the aspired demand. So we need both narrow and wide power concentrations in different applications. Here we provide the new method to control the FWHM in a real-time manner. In the proposed method, we have employed two adjacent focal points to form a new focal point with the desired FWHM. The presented results illustrate that reconfigurable metalens may open the way to various applications in modern communication systems and has a brilliant potential for high-resolution imaging. The presented results illustrate that reconfigurable metalens may open the way to various applications in modern communication systems and has a brilliant potential for high-resolution imaging. The proposed metalens can provide potential applications in integrated nanophotonic devices without conventional limitations.

## 2. Metasurface design

This section presents VO2 in a suitably configured unit cell to realize a 2-bit encoded metasurface. The contribution of four reflection phase digital elements providing 360º reflection phase with a step of 90º is required to mimic the "00". "01", "10'''', and "11" binary states, respectively, depending on the VO2 segments phase. The schematic of the proposed metasurface at the operating frequency of 2.05 THz is depicted in Figure 1. This figure shows the geometry of the suggested three-layer unit cell of gold, silicon, and VO2 from bottom to top. The periodicity of the unit cell is $P = 16$ μm. A gold ground has been fixed in the first layer to prevent wave transmission. The spacer dielectric substrate is silicon with a relative permittivity of $\varepsilon_r = 11.9$ and $\tan\delta = 0.00025$. Also, the thickness of the embedded spacer dielectric is $h = 28$ μm. Lastly, three VO2 segments are designated on the silicon layer with a thickness of 1 μm to fulfill the reflection phase change. After carrying out the required optimization, the dimensional parameters of the structure, shown in Figure 1 (b), have been chosen as $a_1 = 16$ μm, $a_2 = 8$ μm, $a_3 = 3.6$ μm, $a_4 = 2$ μm, and $a_5 = 4$ μm. The employed VO2 segments in this paper have fixed size but different conductivity. Therefore, each unit cell identical geometry allows us to call them bias-encoded unit cell, not geometry-encoded ones. As can be observed schematically in Figure 1, an external reprogrammable device can be employed to deliver the desired bias voltages at its output and controls each segment temperature to set the operational status of digital unit cells individually. By a real-time switch between arbitrary digital sequences, a different focal point can be obtained. In the unit cell simulations, periodic boundary conditions are applied in both x and y directions to consider the mutual coupling influence between adjacent elements and simulate infinite arrangement in both directions. Additionally, the floquet ports are assigned to the z-direction to illuminate incident y-polarized waves. All simulations are carried out by utilizing CST Microwave Studio with a frequency-domain solver. The reflection amplitude and phase of the 2-bit coding metasurfaces are illustrated in Figure 2. After obtaining unit cells' response under different temperature conditions, the complete metasurface is modeled as an array of such unit cells with a predetermined encoded phase map.

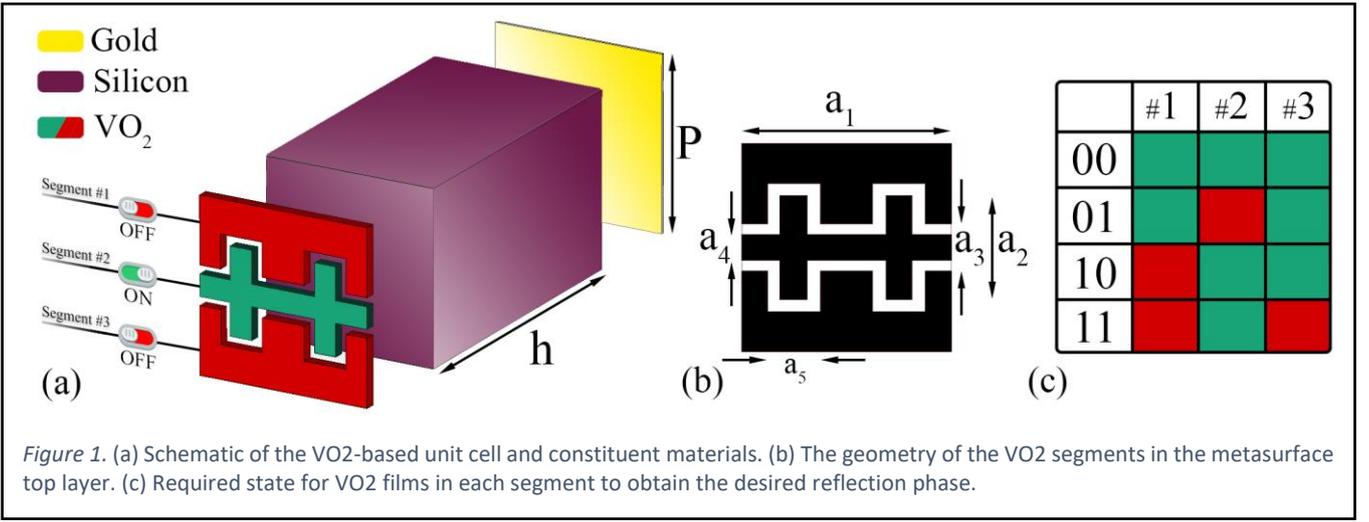

*Figure 1.* (a) Schematic of the VO2-based unit cell and constituent materials. (b) The geometry of the VO2 segments in the metasurface top layer. (c) Required state for VO2 films in each segment to obtain the desired reflection phase.

There are several THz reconfigurable devices that employ VO2 films or pads as thermal- or electrical-sensitive elements [34]. In the metalens design, VO2 particles' conductivity in each unit cell segment is varied to tailor the required reflection phase. At the same time, the configuration of the metasurface is kept constant with predetermined geometry. VO2 relative permittivity is about 9 in the insulating phase, while the conductivity is smaller than 200 S/m [35]. Once the VO2 passes transition temperature, it changes to the metallic state. In this case, relative permittivity is constant, but the conductivity changes by approximately five orders of magnitude during this transition [34]. One approach for basing the VO2-based structures is Joule-heat-induced phase transition actuation caused by the current in VO2 films [36]. The trigger for phase transition depends on the carrier injection or the Poole-Frenkel field effect to produce the electrical current, and the Joule heat generated by the electrical current is sufficient for thermal-induced phase transition before the carrier concentration reaches the critical value for the Mott phase transition [37]. In addition to the methods mentioned above, optical stimulation is a proper method for controlling metal-insulator characteristics [38]. Unlike the electrical stimulation method, structure response time is not linked to geometrical electric parameters. In general, the outcome influenced by the thermal and electrical stimulations is mainly the thermal effect. Therefore, the switching time between two digital states for each VO2 segment in the metasurface cannot be as fast as optical stimulation. The switching time for this technique is expected to be on the orders of sub-seconds. Regarding the proposed metasurface's potential fabrication process, we designated two opposite electrodes and a ground plane as the negative electrode and three pieces of separated patches as the positive electrodes. Adding these new substrates and a metallic patch underneath the structure does not influence the unit cell's reflection response. A resistive heater is used here, which causes the temperature to reach the required temperature for a phase transition. To local biasing of each VO2 patches independently, two thin Au patches are placed on each VO2 layer. Three metallic via holes are drilled through the negative electrode and one metallic via hole on each positive electrode to connect these electrodes to VO2 layers. Now a DC current is applied to the resistive heater on the structure, so the phase transition temperature can be obtained by controlling an external voltage source for each VO2 layer independently

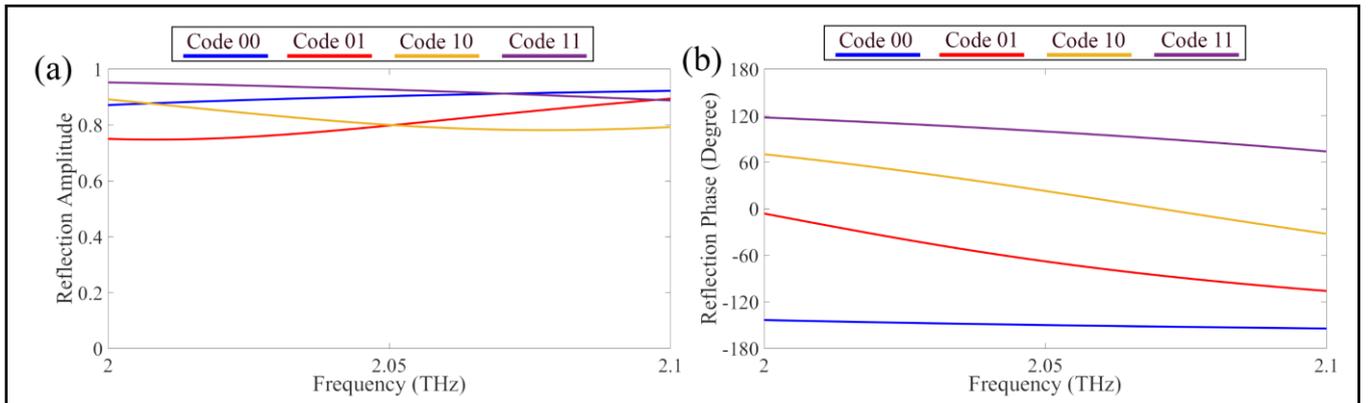

*Figure 2.* Broadband performance of the (a) amplitude and (b) phase of the reflected beam under y-polarized illumination for the suggested VO2-based reconfigurable coding metasurface from 2 to 2.1 THz, respectively. According to extracted results, 2.05 THz is an operating frequency in this work.

## 3. Principle of Multi Focus Method

Huygen's theory can generate specific electromagnetic responses under incident waves illumination. The electromagnetic manipulation is produced by the induced surface electric and magnetic current, which can be scattered electric and magnetic field distribution above the metasurface indirectly according to the boundary conditions. We consider each unit cell as a secondary point source, and the incident illumination can be concentrated on defined positions by adding extra abrupt phase changes to the reflected wave utilizing the suggested unit cell.

In recent years the various method has been presented to design multifocal metalens. In the first methodology, the metasurface is divided into many segments, and each segment concentrates the beam on the desired point [21], [39]–[41]. Metasurface division can be done longitudinally or radially, but this method has several disadvantages. For instance, in this method, metasurface can not control the power ratio in the concentration points. Also, we are not able to concentrate the several focal points in the same region or in the same axis normal to metasurface. In other novel studies, the authors mentioned a new method to achieve multifocal metalens by controlling both the amplitude and phase of the fundamental unit cells [42]–[44]. This method is the most accurate method to obtain the multifocal point in the desired spots of the space. Nonetheless, designing the tunable unit cell, which can control amplitude and phase simultaneously, is a difficult step in this method. In the next step, several papers have been published that using phase-only metasurface to control power distribution on the focal points [45]–[48]. This method is a simplified version of the previous method and has the brilliant capability to control the power ratio on the different focal points.

The holographic algorithm calculates phase layout over the metasurface to obtain the desired electric field distribution [49]. The approach to realize a multifocal pattern is based on selecting ideal point sources as virtual sources at the predetermined focal points [46]. The schematic representation for this approach is illustrated in Figure 3. Then, by superposing a radiated field generated by virtual sources, the virtual electric field distribution can be calculated. Here the electromagnetic field propagation from each virtual source is represented by the Green function [45]. Then the virtual electric field can be mapped onto the recording plane where the metamirror is placed at the plane of $z = 0$ ($\varphi(x, y)$). Because the calculation of phase conjugation is equivalent to the time-reversal for the incident monochromatic wave, the phase delay of each unit cell should be set as $-\varphi(x, y)$ to generate a multifocal image in the desired plane [45]. Consequently, the reconstructed electric field distribution is concentrated on the predesigned focal points. In the general case, the electric field intensity decreases linearly with the reciprocal of the distance between the unit cell and focal points [22]. For a multi-focus scenario, the intensity of focal points could be adjusted with arbitrary weights. The desired reflection phase at the position of the different elementary unit cell is determined by [45]

$$-\phi(x_k, y_k) = -\arg\left\{\sum_{i=0}^{N} w_i G(r_{m,i})\right\} = -\arg\left\{\sum_{i=0}^{N} w_i \frac{e^{-jk_0 r_{m,i}}}{4\pi r_{m,i}}\right\} \quad (k = 1 \text{ to } M), \tag{1}$$

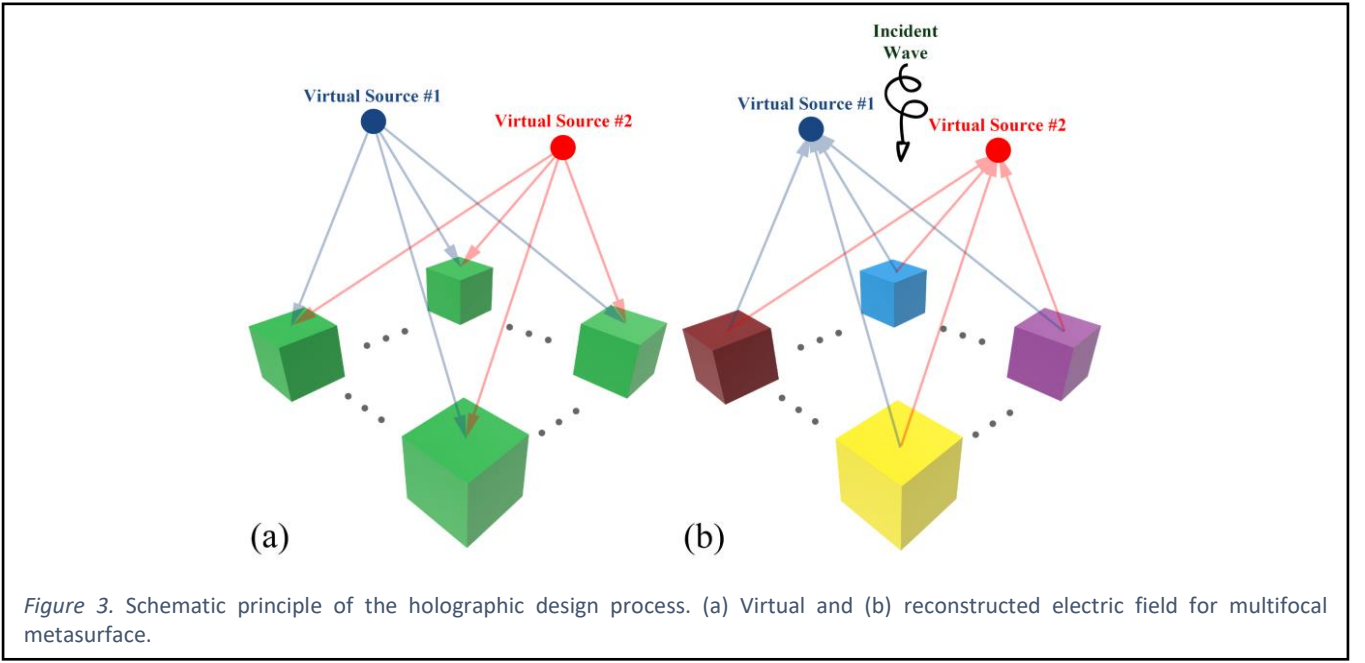

*Figure 3.* Schematic principle of the holographic design process. (a) Virtual and (b) reconstructed electric field for multifocal metasurface.

where $r_{m,i} = \sqrt{(x_k - x_i)^2 + (y_k - y_i)^2 + z_i^2}$ is the distance between the $k$-th unit cell located at $(x_k, y_k, 0)$ ($k = 1$ to $M$) and the $i$-th focal point at $(x_i, y_i, z_i)$ ($i = 1$ to $N$). Also, $k_0$ is the propagation constant, $w_i$ is the weight factor of $i$-th virtual source, which presents the intensity ratio of the $i$-th focal point to the other ones. Notably, we assumed the maximum value for $w_i$ is set to 1. Then, the reflection phase is determined to be $-\varphi(x_k, y_k, 0)$ for the element located at $(x_k, y_k, 0)$.

Next, based on the established reflection phase distribution, the reconstructed electric field distribution can be calculated through the superposition of the electric field component generated by each unit cell. Since the electric field component generated by each secondary radiator can be described by the Green function theoretically, the reconstructed electric field distribution under the incident plane wave is described as

$$E(x, y, z) = \sum_{k=1}^{M} e^{jk_0 \phi(x_k, y_k, 0)} G(r_{m,k}) = \sum_{k=1}^{M} \frac{e^{j(\arg\{\sum_{i=1}^{n} w_i \frac{e^{-jk_0 r_{m,i}}}{4\pi r_{m,i}}\} - k_0 r_{m,k})}}{4\pi r_{m,k}} \quad (2)$$

where $r_{m,k} = \sqrt{(x_k - x)^2 + (y_k - y)^2 + z^2}$ denotes the distance between the unit cell at $(x_k, y_k, 0)$ and the observation point at $(x, y, z)$ on the image plane. Also, in the above equation $|E(x, y, z)|^2$ denotes the reflected power on the reconstructed image plane.

## 4. Results and Discussion

### 4.1 Multi-Focal Points

This section investigates the proposed metasurface's capability in focusing the electromagnetic wave at the operating frequency of 2.05 THz. For instance, we calculated the required coding layout of building blocks with different phase distribution to concentrate reflected energy on the desired points. The presented metasurface in this work is composed of 60 × 60 particles, which is equal to 6.4λ × 6.4λ at the operating frequency. It is worth mentioning that the proposed metasurface, unlike our previous work [21], can build multifocal points in different locations simultaneously, without dividing a structure into multi-segment parts that each segment generates one focal point in space. So we have fewer restrictions on this method, and we can utilize this method for the general case. Also, compared to other research by one of the authors [22], the efficiency is increased in this research due to lower loss in VO2 in contrast to graphene monolayer. Consequently, VO2 is a promising candidate for low loss wave manipulation in the THz regime compared to other tunable and phase change materials. In order to evaluate the performance of the VO2-based designed metalens, we have proposed four different scenarios for the focal points.

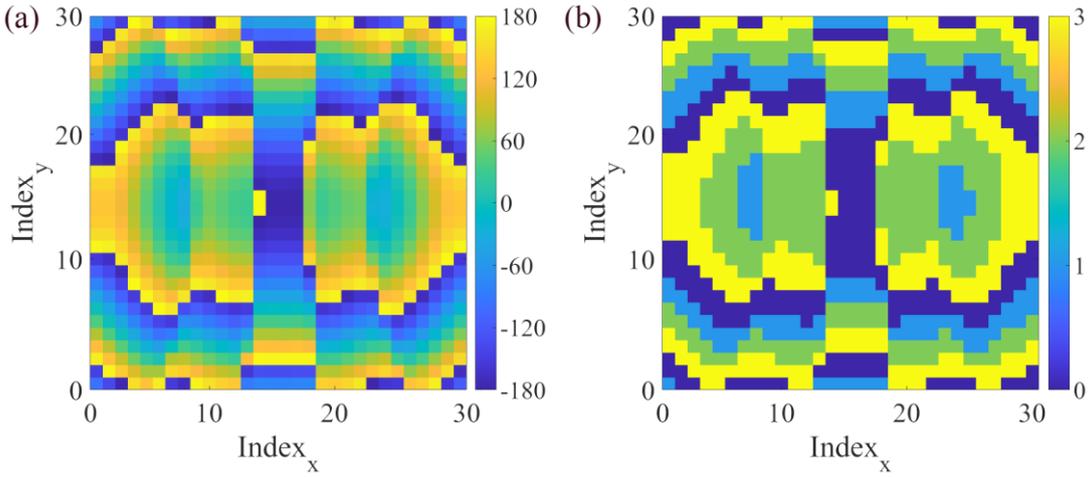

*Figure 4.* (a) Continuous and (b) 2-bit encoded reflection spatial phase layout to concentrate the beam at (250 μm, 0 μm, 400 μm) and (-250 μm, 0 μm, 400 μm) with an equal weight factor of unit.

Figure 4 shows a metasurface continuous and 2-bit encoded reflection phase profile to concentrate the beam at (250 μm, 0 μm, 400 μm) and (−250 μm, 0 μm, 400 μm) with an equal weight factor of unit. In this case, the unit weight factor shows that the same amount of energy is divided between focal points. Simulated results for electric field distribution have been indicated in Figure 5. As we can see in Figure 5 (a), the metasurface produces efficient electric field concentration very close to the desired points. As depicted in the three-dimensional results in Figures 5 (b), side lobes appear around the focal spot due to the planar metasurface's finite and discontinuity in coding metasurface. Also, side lobes level value and focal point location accuracy can be improved with larger metasurfaces or smaller unit cell dimensions [22]. To evaluate the simulation results, we have presented theoretical results in Figure 6 based on the analytical method that is demonstrated in [50], [51] for electric field calculation. In order to elucidate the performance of the metalens, we extract the important lens parameter. The full width at half-maximum of the focal point in the x−y plane is 0.6λ in the operating frequency, which is near the diffraction limit. The unit cell's dimension compares to operating wavelength is an essential parameter in metalens design. In our study, the VO2-based unit cell is 0.11 of operating wavelength, so we can declare that element size is excellent enough to have a good focusing resolution as well as we can consider the electromagnetic coupling between particles. Furthermore, the size of a metalens is another important parameter in the design procedure. It is also considered to make sure that the whole metasurface will have enough dimension to yield a one complete phase range of 360º. By following these simple rules, we can design a high efficient metalens which can generate a proper concentration spot at the desired point.

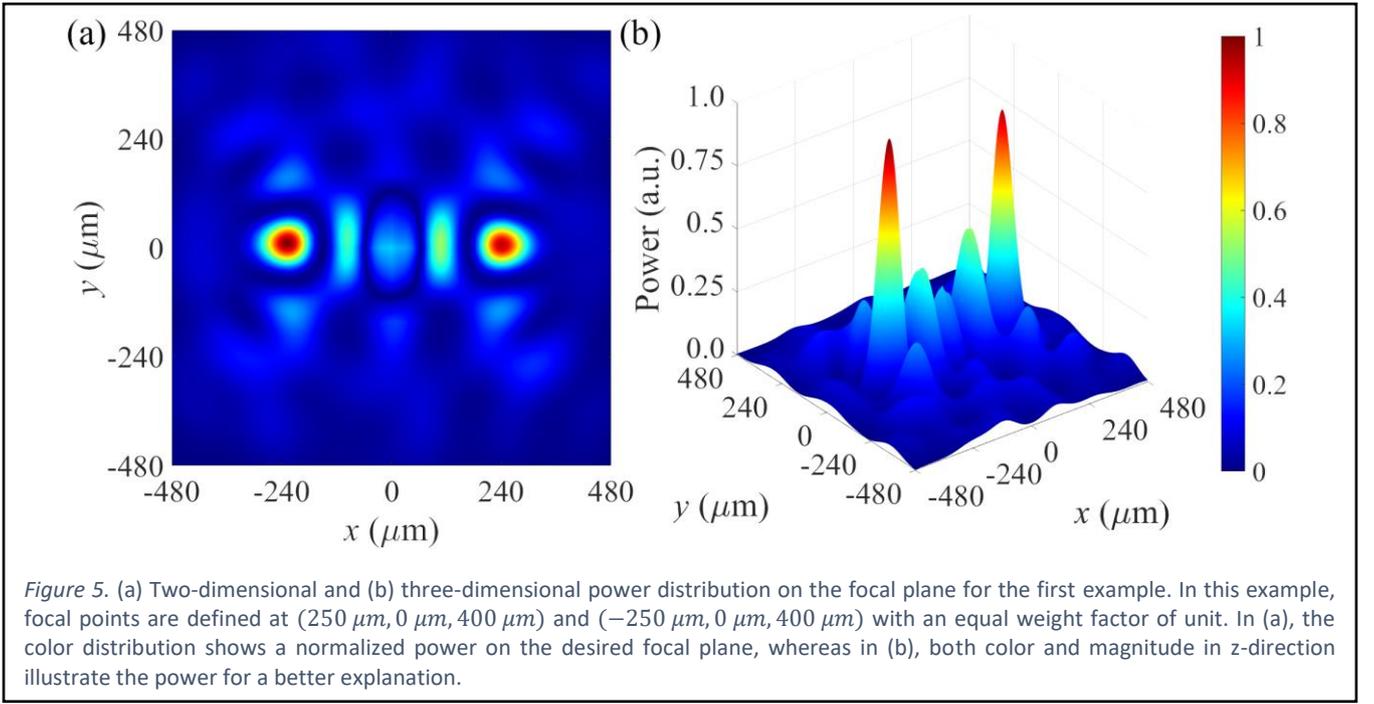

*Figure 5.* (a) Two-dimensional and (b) three-dimensional power distribution on the focal plane for the first example. In this example, focal points are defined at $(250\ \mu m, 0\ \mu m, 400\ \mu m)$ and $(-250\ \mu m, 0\ \mu m, 400\ \mu m)$ with an equal weight factor of unit. In (a), the color distribution shows a normalized power on the desired focal plane, whereas in (b), both color and magnitude in z-direction illustrate the power for a better explanation.

The next example is two symmetric points with the same position, which is mentioned in the first example but with different weight factors. In this example, $w_1 = 1$ and $w_2 = 0.7$, so the power concentrated at the second point is about half the power concentrated at the first point. The simulation's extracted results are illustrated in Supplementary Figure S1. This example demonstrates that we can control power density at the desired point by choosing an appropriate value for weight factors. In this example, symmetric focal points are selected, but this method for controlling the power is not limited to the symmetric focal points and can be used in asymmetric focal points missions.

The third example is an important example that overcomes the limitation in our previous method (See ref. [21]) and investigates the new method's advantage. In this case, the metasurface concentrates energy in two different focal points, located in the same axis perpendicular to the metasurface. The selected focal points are $(0\ \mu m, 0\ \mu m, 300\ \mu m)$, and $(0\ \mu m, 0\ \mu m, 600\ \mu m)$, which are located in reasonable distance on the z-axis. Also, both focal points' weight factor is unit, which shows that dedicated power for focal points is alike. It is worth noting that the equal weight factor does not imply that the concentrated power in focal points is equal. Indeed, it shows that the energy allocated as a secondary source input is equal. Also, the focal point distance from the metasurface and focal point position in the half-space are other important parameters that can affect concentrated power. The calculated results in the two different focal planes are illustrated in Supplementary

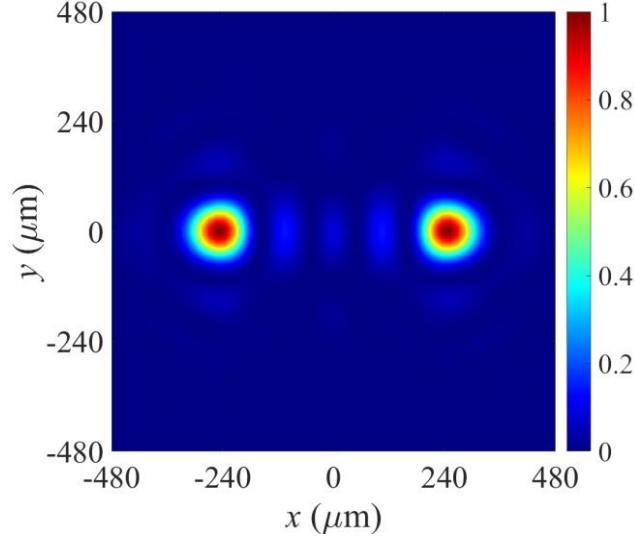

*Figure 6.* Theoretically calculated power distribution on the focal plane for the first example. In this example, focal points defined at $(250\ \mu m, 0\ \mu m, 400\ \mu m)$ and $(-250\ \mu m, 0\ \mu m, 400\ \mu m)$ with an equal weight factor of unit.

Figure S2. These two-dimensional results show that the reflected energy is concentrated in the focal plane's center, as we predicted before. The first focal point has a narrower half-power bandwidth because it is closer to the metasurface, whereas the second concentration region is wider because it is farther from the surface. Thus, the farther we go from the metasurface, the beam's diameter increases, and energy concentration accuracy decreases.

The last example introduces another feature of the reconfigurable metasurface, which is manipulation of electromagnetic waves at more than two focal spots. Four focal points are determined here, designated symmetrically in each quarter of half-space with an equal weight factor. The position of each focal points is as follow $(300\ \mu m, 300\ \mu m, 300\ \mu m)$, $(-300\ \mu m, 300\ \mu m, 300\ \mu m)$, $(-300\ \mu m, -300\ \mu m, 300\ \mu m)$, and $(300\ \mu m, -300\ \mu m, 300\ \mu m)$. The schematic of 2-bit encoded phase profiles is depicted in Supplementary Figure S3, and eventually, Supplementary Figure S4 confirmed the simulation results with theoretical results in a perfect way. The final example shows that the proposed structure can concentrate reflected energy in multifocal points, which has a brilliant potential for multi-channel communication and imaging.

*4.2 Extended FWHM and Asymmetric Power Distribution*

In this section, we will investigate the capability of the multifocal lens to extend the FWHM. The FWHM of a defined focal point can be increased by placing two focal points close to each other. In order to compare the method results, we start our analysis by investigating one focal point characteristic. So, we assume that we want to concentrate the reflected beam on one focal spot $(0\ \mu m, 0\ \mu m, 400\ \mu m)$. We have used Equation (1) to obtain the required phase arrangement, whereas in this equation, $N = 1$. As shown in Figure 7 (a), the two-dimensional power distribution shows a circular concentration point in the focal plane. Additionally, Figure 7 (d) illustrates the one-dimensional result in the determined curve. According to extracted results based on full-wave simulation, FWHM is 80.2 μm. In the next case, we considered two different focal points with the same weight factor symmetrical about the mentioned single focal point. These two focal points are located in the observation plane at $(30\ \mu m, 0\ \mu m, 400\ \mu m)$ and $(-30\ \mu m, 0\ \mu m, 400\ \mu m)$. The power distribution illustrated in Figure 7 (b) shows that the concentration waist increased in x-direction compared to the last example. Indeed, because the distance between two focal points is less than the FWHM in the single-focus example, the focal points had been merged and composed a wider concentration region. Based on the simulation results in Figure 7 (d), FWHM is increased to 123.2 μm, and a new single focus with extended width is constructed in z-plane is observed. According to the declared results in this section, we can find that the FWHM is developed by more than 50 percent. The most important point in this method is choosing the right focal points. To demonstrate this point well, we have provided another example in this section. In this instance, we have increased the distance between the focal point. Thus, we assume focal points are located at $(90\ \mu m, 0\ \mu m, 400\ \mu m)$ and $(-90\ \mu m, 0\ \mu m, 400\ \mu m)$, and they are farther away from the FWHM value in the single-focus example. The two-dimensional and one-dimensional normalized power distribution is exhibited in Figure 7 (a) and Figure 7 (d), respectively. It is clear that the two focal points are far apart, so the maximum power spots are separated, and we can not consider these regions as one compact concentration point. This example demonstrates that the focal points' position is an indispensable parameter in this method, and designers have to choose them very carefully.

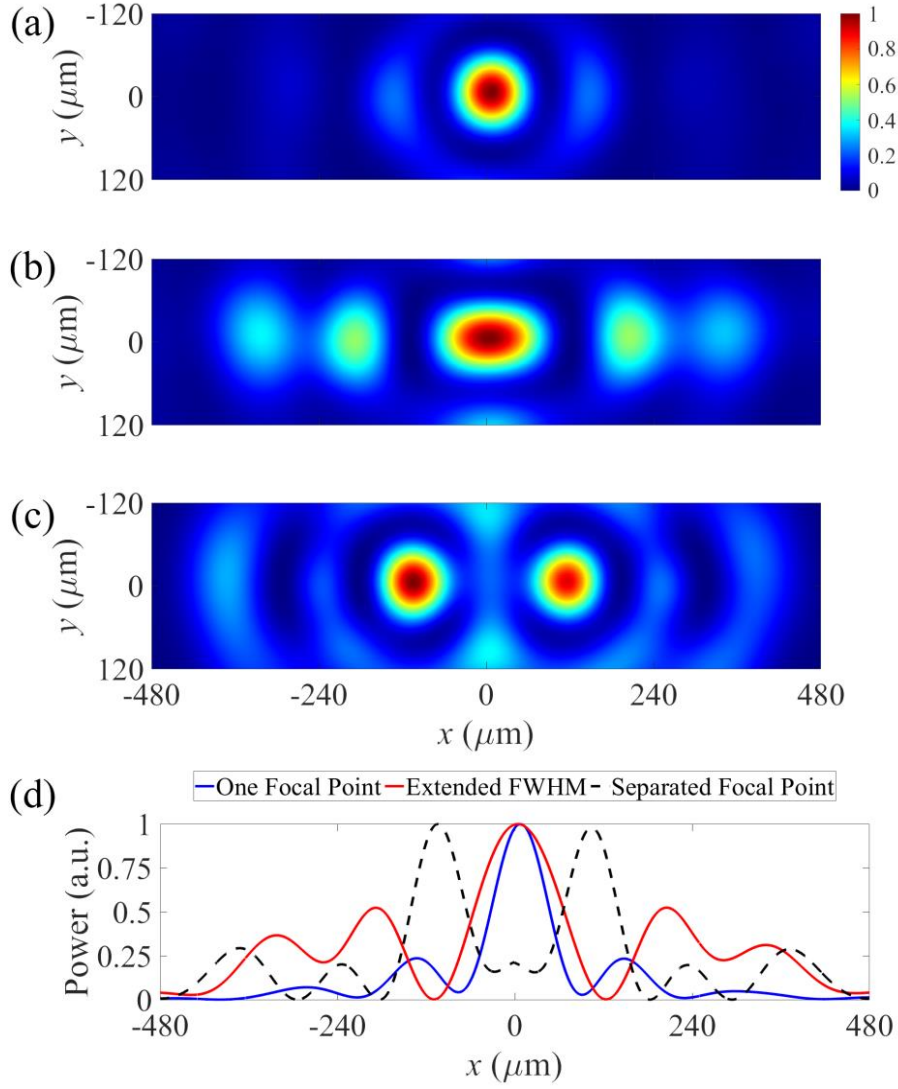

*Figure 7.* (a-c) Two-dimensional normalized power distribution on the focal plane for the three examples. (a) One focal point at $(0\ \mu m, 0\ \mu m, 400\ \mu m)$. (b) Two adjacent focal points at $(30\ \mu m, 0\ \mu m, 400\ \mu m)$ and $(-30\ \mu m, 0\ \mu m, 400\ \mu m)$. (b) Two separate focal points at $(90\ \mu m, 0\ \mu m, 400\ \mu m)$ and $(-90\ \mu m, 0\ \mu m, 400\ \mu m)$. (d) Corresponding one-dimensional normalized power distribution for the proposed examples.

This section has only mentioned simple examples to extend FWHM in the x-direction, whereas designers can tune the desired parameter by changing an encoded layout through the metasurface. For instance, we can put another symmetric focal point in the y-direction to control FWHM in this direction in a real-time manner. In addition to the simple example provided in this section, designers can employ more than two focal points and increase it to multifocal points in the specific direction to extend the FWHM in the desired direction or control it dynamically. As the last application, we can assign unequal values for weight factors to achieve asymmetric power distribution, which can have various applications in complex imaging methods.

**Conclusion**

In conclusion, a general paradigm to design reflection-type metasurface with reconfigurable functionalities is proposed in this paper. We have employed distinct VO2-based metasurface to realize multifunctional metalenses that produce any arbitrary focal points with a high-resolution concentration in space at the operating frequency of 2.05 THz. The designed unit cell has a unique capability to switch between four digital states, "00", "01", "10" and "11" by controlling VO2 material in an insulating or fully metallic phase. In addition, we have proposed a new method to control the FWHM in the metalens by employing two neighbor focal points in space. The full-wave simulation shows that all designed metalenses agree very well with the theoretical expectations and have satisfying focusing characteristics. Moreover, each focal point's power intensity can be tuned asymmetrically by adjusting the weight factor in spatial phase distribution. These results show that the superposition method is a practical approach and has potential applications in designing integrated THz devices on the sub-wavelength scale, micro-manipulating, telecommunications, and parallel-imaging technology.

**References**

[1]     K. Zhang, Y. Yuan, X. Ding, B. Ratni, S. N. Burokur, and Q. Wu, "High-Efficiency Metalenses with Switchable Functionalities in Microwave Region," *ACS Appl. Mater. Interfaces*, vol. 11, no. 31, pp. 28423–28430, Aug. 2019, doi: 10.1021/acsami.9b07102.

[2]     S. Fallah, K. Rouhi, and A. Abdolali, "Optimized chemical potential graphene-based coding metasurface approach for dynamic manipulation of terahertz wavefront," *J. Phys. D. Appl. Phys.*, vol. 53, no. 8, p. 085102, Feb. 2020, doi: 10.1088/1361-6463/ab572f.

[3]     A. Abdolali, A. Momeni, H. Rajabalipanah, and K. Achouri, "Parallel integro-differential equation solving via multi-channel reciprocal bianisotropic metasurface augmented by normal susceptibilities," *New J. Phys.*, vol. 21, no. 11, p. 113048, Nov. 2019, doi: 10.1088/1367-2630/ab26f8.

[4]     A. Momeni, H. Rajabalipanah, A. Abdolali, and K. Achouri, "Generalized Optical Signal Processing Based on Multioperator Metasurfaces Synthesized by Susceptibility Tensors," *Phys. Rev. Appl.*, vol. 11, no. 6, p. 064042, Jun. 2019, doi: 10.1103/PhysRevApplied.11.064042.

[5] A. Babaee, A. Momeni, A. Abdolali, and R. Fleury, "Parallel Optical Computing Based on MIMO Metasurface Processors with Asymmetric Optical Response," *arXiv*, Apr. 2020, [Online]. Available: http://arxiv.org/abs/2004.02948.

[6] M. Rahmanzadeh, H. Rajabalipanah, and A. Abdolali, "Analytical Investigation of Ultrabroadband Plasma–Graphene Radar Absorbing Structures," *IEEE Trans. Plasma Sci.*, vol. 45, no. 6, pp. 945–954, Jun. 2017, doi: 10.1109/TPS.2017.2700724.

[7] M. Mohammadi, H. Rajabalipanah, and A. Abdolali, "A theoretical investigation on reciprocity-inspired wide-angle spectrally-selective THz absorbers augmented by anisotropic metamaterials," *Sci. Rep.*, vol. 10, no. 1, p. 10396, Dec. 2020, doi: 10.1038/s41598-020-67399-3.

[8] K. Rouhi, A. Abdolali, and S. Fallah, "Wideband THz Low-Scattering Surface Based on Combination of Diffusion and Absorption," *Arxiv*, pp. 1–16, Jan. 2020, [Online]. Available: http://arxiv.org/abs/2001.09289.

[9] M. Rahmanzadeh and A. Khavasi, "Perfect anomalous reflection using a compound metallic metagrating," *Opt. Express*, vol. 28, no. 11, p. 16439, May 2020, doi: 10.1364/OE.393137.

[10] A. Momeni, K. Rouhi, H. Rajabalipanah, and A. Abdolali, "An Information Theory-Inspired Strategy for Design of Re-programmable Encrypted Graphene-based Coding Metasurfaces at Terahertz Frequencies," *Sci. Rep.*, vol. 8, no. 1, p. 6200, Dec. 2018, doi: 10.1038/s41598-018-24553-2.

[11] P. R. West *et al.*, "All-dielectric subwavelength metasurface focusing lens," *Opt. Express*, 2014, doi: 10.1364/oe.22.026212.

[12] M. Khorasaninejad *et al.*, "Achromatic Metasurface Lens at Telecommunication Wavelengths," *Nano Lett.*, vol. 15, no. 8, pp. 5358–5362, Aug. 2015, doi: 10.1021/acs.nanolett.5b01727.

[13] M. Khorasaninejad and F. Capasso, "Metalenses: Versatile multifunctional photonic components," *Science (80-. ).*, vol. 358, no. 6367, p. eaam8100, Dec. 2017, doi: 10.1126/science.aam8100.

[14] M. L. Tseng *et al.*, "Metalenses: Advances and Applications," *Adv. Opt. Mater.*, vol. 6, no. 18, p. 1800554, Sep. 2018, doi: 10.1002/adom.201800554.

[15] Q. Yang *et al.*, "Efficient flat metasurface lens for terahertz imaging," *Opt. Express*, vol. 22, no. 21, p. 25931, Oct. 2014, doi: 10.1364/OE.22.025931.

[16] C. Schlickriede, S. S. Kruk, L. Wang, B. Sain, Y. Kivshar, and T. Zentgraf, "Nonlinear Imaging with All-Dielectric Metasurfaces," *Nano Lett.*, vol. 20, no. 6, pp. 4370–4376, Jun. 2020, doi: 10.1021/acs.nanolett.0c01105.

[17] P. Yang, R. Yang, and J. Zhu, "Wave manipulation with metasurface lens in the cassegrain system," *J. Phys. D. Appl. Phys.*, vol. 52, no. 35, p. 355101, Aug. 2019, doi: 10.1088/1361-6463/ab27a7.

[18] Y. Li *et al.*, "Broadband unidirectional cloaks based on flat metasurface focusing lenses," *J. Phys. D. Appl. Phys.*, vol. 48, no. 33, p. 335101, Aug. 2015, doi: 10.1088/0022-3727/48/33/335101.

[19] H. Li, G. Wang, T. Cai, J. Liang, and H. Hou, "Bifunctional circularly-polarized lenses with simultaneous geometrical and propagating phase control metasurfaces," *J. Phys. D. Appl. Phys.*, vol. 52, no. 46, p. 465105, Nov. 2019, doi: 10.1088/1361-6463/ab39ac.

[20] M. Park, C. Park, Y. S. Hwang, E. Kim, D. Choi, and S. Lee, "Virtual-Moving Metalens Array Enabling Light-Field Imaging with Enhanced Resolution," *Adv. Opt. Mater.*, p. 2000820, Sep. 2020, doi: 10.1002/adom.202000820.

[21] R. Kargar, K. Rouhi, and A. Abdolali, "Reprogrammable multifocal THz metalens based on metal–insulator transition of VO2-

assisted digital metasurface," *Opt. Commun.*, vol. 462, p. 125331, May 2020, doi: 10.1016/j.optcom.2020.125331.

[22]  S. E. Hosseininejad *et al.*, "Reprogrammable Graphene-based Metasurface Mirror with Adaptive Focal Point for THz Imaging," *Sci. Rep.*, vol. 9, no. 1, p. 2868, Dec. 2019, doi: 10.1038/s41598-019-39266-3.

[23]  A. Komar *et al.*, "Dynamic Beam Switching by Liquid Crystal Tunable Dielectric Metasurfaces," *ACS Photonics*, 2018, doi: 10.1021/acsphotonics.7b01343.

[24]  Q. Wang *et al.*, "Optically reconfigurable metasurfaces and photonic devices based on phase change materials," *Nat. Photonics*, 2016, doi: 10.1038/nphoton.2015.247.

[25]  H. Barati Sedeh, M. M. Salary, and H. Mosallaei, "Time-varying optical vortices enabled by time-modulated metasurfaces," *Nanophotonics*, vol. 9, no. 9, pp. 2957–2976, Jul. 2020, doi: 10.1515/nanoph-2020-0202.

[26]  K. Rouhi, H. Rajabalipanah, and A. Abdolali, "Real-Time and Broadband Terahertz Wave Scattering Manipulation via Polarization-Insensitive Conformal Graphene-Based Coding Metasurfaces," *Ann. Phys.*, vol. 530, no. 4, p. 1700310, Apr. 2018, doi: 10.1002/andp.201700310.

[27]  K. Rouhi, H. Rajabalipanah, and A. Abdolali, "Multi-bit graphene-based bias-encoded metasurfaces for real-time terahertz wavefront shaping: From controllable orbital angular momentum generation toward arbitrary beam tailoring," *Carbon N. Y.*, vol. 149, pp. 125–138, Aug. 2019, doi: 10.1016/j.carbon.2019.04.034.

[28]  S. E. Hosseininejad, K. Rouhi, M. Neshat, A. Cabellos-Aparicio, S. Abadal, and E. Alarcon, "Digital Metasurface Based on Graphene: An Application to Beam Steering in Terahertz Plasmonic Antennas," *IEEE Trans. Nanotechnol.*, vol. 18, pp. 734–746, 2019, doi: 10.1109/TNANO.2019.2923727.

[29]  M. Kiani, M. Tayarani, A. Momeni, H. Rajabalipanah, and A. Abdolali, "Self-biased tri-state power-multiplexed digital metasurface operating at microwave frequencies," *Opt. Express*, vol. 28, no. 4, p. 5410, Feb. 2020, doi: 10.1364/OE.385524.

[30]  M. Kiani, A. Momeni, M. Tayarani, and C. Ding, "Spatial Wave Control Using a Self-biased Nonlinear Metasurface at Microwave Frequencies," *arXiv*, Jul. 2020, [Online]. Available: http://arxiv.org/abs/2007.06480.

[31]  M. A. Baqir and P. K. Choudhury, "On the VO 2 metasurface-based temperature sensor," *J. Opt. Soc. Am. B*, vol. 36, no. 8, p. F123, Aug. 2019, doi: 10.1364/JOSAB.36.00F123.

[32]  Y. Chen *et al.*, "Free-standing SWNTs/VO 2 /Mica hierarchical films for high-performance thermochromic devices," *Nano Energy*, vol. 31, pp. 144–151, Jan. 2017, doi: 10.1016/j.nanoen.2016.11.030.

[33]  P. Guo *et al.*, "Conformal Coating of a Phase Change Material on Ordered Plasmonic Nanorod Arrays for Broadband All-Optical Switching," *ACS Nano*, vol. 11, no. 1, pp. 693–701, Jan. 2017, doi: 10.1021/acsnano.6b07042.

[34]  C. Zhang *et al.*, "Active Control of Terahertz Waves Using Vanadium-Dioxide-Embedded Metamaterials," *Phys. Rev. Appl.*, vol. 11, no. 5, p. 054016, May 2019, doi: 10.1103/PhysRevApplied.11.054016.

[35]  F. Ding, S. Zhong, and S. I. Bozhevolnyi, "Vanadium Dioxide Integrated Metasurfaces with Switchable Functionalities at Terahertz Frequencies," *Adv. Opt. Mater.*, vol. 6, no. 9, p. 1701204, May 2018, doi: 10.1002/adom.201701204.

[36]  G. Gopalakrishnan, D. Ruzmetov, and S. Ramanathan, "On the triggering mechanism for the metal–insulator transition in thin film VO2 devices: electric field versus thermal effects," *J. Mater. Sci.*, vol. 44, no. 19, pp. 5345–5353, Oct. 2009, doi: 10.1007/s10853-009-3442-7.

[37] S. Kumar, M. D. Pickett, J. P. Strachan, G. Gibson, Y. Nishi, and R. S. Williams, "Local Temperature Redistribution and Structural Transition During Joule-Heating-Driven Conductance Switching in VO 2," *Adv. Mater.*, vol. 25, no. 42, pp. 6128–6132, Nov. 2013, doi: 10.1002/adma.201302046.

[38] J. Gu *et al.*, "Active control of electromagnetically induced transparency analogue in terahertz metamaterials," *Nat. Commun.*, vol. 3, no. 1, p. 1151, Jan. 2012, doi: 10.1038/ncomms2153.

[39] X. Chen *et al.*, "Longitudinal Multifoci Metalens for Circularly Polarized Light," *Adv. Opt. Mater.*, vol. 3, no. 9, pp. 1201–1206, Sep. 2015, doi: 10.1002/adom.201500110.

[40] Q. Chen *et al.*, "High numerical aperture multifocal metalens based on Pancharatnam–Berry phase optical elements," *Appl. Opt.*, vol. 57, no. 27, p. 7891, Sep. 2018, doi: 10.1364/AO.57.007891.

[41] F. Ding, Y. Chen, and S. I. Bozhevolnyi, "Gap-surface plasmon metasurfaces for linear-polarization conversion, focusing, and beam splitting," *Photonics Res.*, vol. 8, no. 5, p. 707, May 2020, doi: 10.1364/PRJ.386655.

[42] M. Hashemi, A. Moazami, M. Naserpour, and C. J. Zapata-Rodríguez, "A broadband multifocal metalens in the terahertz frequency range," *Opt. Commun.*, vol. 370, pp. 306–310, Jul. 2016, doi: 10.1016/j.optcom.2016.03.031.

[43] X. Wang, J. Ding, B. Zheng, S. An, G. Zhai, and H. Zhang, "Simultaneous Realization of Anomalous Reflection and Transmission at Two Frequencies using Bi-functional Metasurfaces," *Sci. Rep.*, vol. 8, no. 1, p. 1876, Dec. 2018, doi: 10.1038/s41598-018-20315-2.

[44] G. Ding *et al.*, "Direct routing of intensity-editable multi-beams by dual geometric phase interference in metasurface," *Nanophotonics*, vol. 9, no. 9, pp. 2977–2987, Jun. 2020, doi: 10.1515/nanoph-2020-0203.

[45] Z. Wang, X. Ding, K. Zhang, and Q. Wu, "Spacial Energy Distribution Manipulation with Multi-focus Huygens Metamirror," *Sci. Rep.*, vol. 7, no. 1, p. 9081, Dec. 2017, doi: 10.1038/s41598-017-09474-w.

[46] Y. Wang *et al.*, "Multi-focus hologram utilizing Pancharatnam–Berry phase elements based metamirror," *Opt. Lett.*, vol. 44, no. 9, p. 2189, May 2019, doi: 10.1364/OL.44.002189.

[47] S. Yu, H. Liu, and L. Li, "Design of Near-Field Focused Metasurface for High-Efficient Wireless Power Transfer With Multifocus Characteristics," *IEEE Trans. Ind. Electron.*, vol. 66, no. 5, pp. 3993–4002, May 2019, doi: 10.1109/TIE.2018.2815991.

[48] B. Ratni *et al.*, "Dynamically Controlling Spatial Energy Distribution with a Holographic Metamirror for Adaptive Focusing," *Phys. Rev. Appl.*, vol. 13, no. 3, p. 034006, Mar. 2020, doi: 10.1103/PhysRevApplied.13.034006.

[49] Z. Wang *et al.*, "Huygens Metasurface Holograms with the Modulation of Focal Energy Distribution," *Adv. Opt. Mater.*, vol. 6, no. 12, p. 1800121, Jun. 2018, doi: 10.1002/adom.201800121.

[50] H. Rajabalipanah, K. Rouhi, A. Abdolali, S. Iqbal, L. Zhang, and S. Liu, "Real-time terahertz meta-cryptography using polarization-multiplexed graphene-based computer-generated holograms," *Nanophotonics*, Jun. 2020, doi: 10.1515/nanoph-2020-0110.

[51] H. Rajabalipanah, A. Abdolali, and K. Rouhi, "Reprogrammable Spatiotemporally Modulated Graphene-Based Functional Metasurfaces," *IEEE J. Emerg. Sel. Top. Circuits Syst.*, vol. 10, no. 1, pp. 75–87, Mar. 2020, doi: 10.1109/JETCAS.2020.2972928.

# Multifocal Terahertz Lens with Adjustable Focal Points

Ali Abdolali[1,*], Roya Kargar[1], and Kasra Rouhi[1]

[1] Applied Electromagnetic Laboratory, School of Electrical Engineering, Iran University of Science and Technology, Tehran, 1684613114, Iran

E-mail: abdolali@iust.ac.ir

Contents:

# I. Multi-Focal Points Results

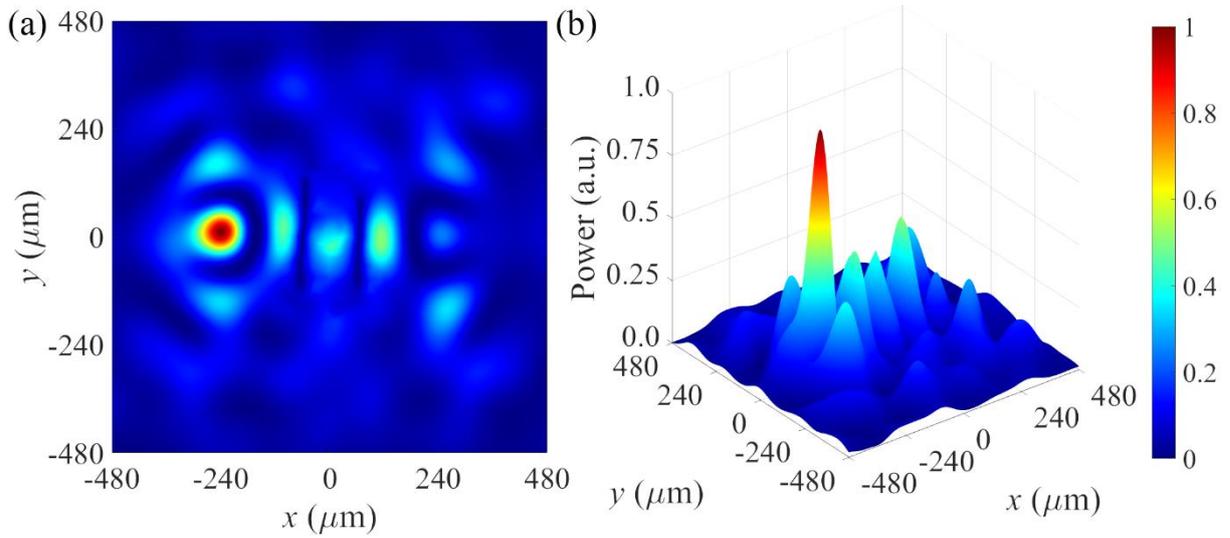

*Supplementary Figure S1* – (a) Two-dimensional and (b) three-dimensional power distribution on the focal plane for the second example. In this example, focal points are defined at $(250\ \mu m, 0\ \mu m, 400\ \mu m)$ and $(-250\ \mu m, 0\ \mu m, 400\ \mu m)$ with an unequal weight factor of $w_1 = 1$ and $w_2 = 0.7$.

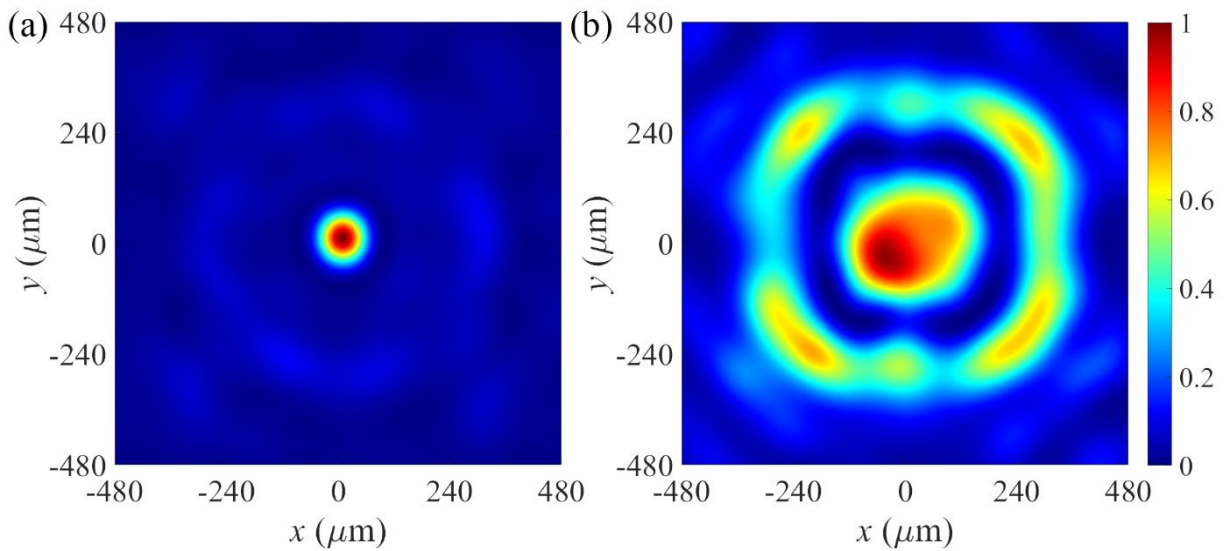

*Supplementary Figure S2* - Two-dimensional power distribution on the two focal planes $z_1 = 300\ \mu m$ and $z_2 = 600\ \mu m$ for the third example. In this example, focal points are defined at $(0\ \mu m, 0\ \mu m, 300\ \mu m)$ and $(0\ \mu m, 0\ \mu m, 600\ \mu m)$ with an equal weight factor of unit.

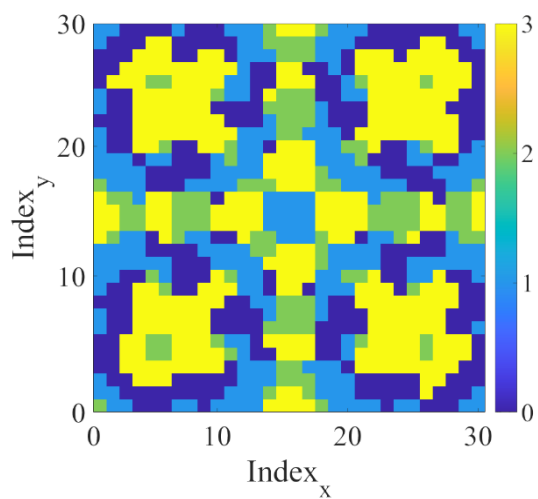

*Supplementary Figure S3* - 2-bit encoded reflection phase layout to concentrate the beam at $(300\ \mu m, 300\ \mu m, 300\ \mu m)$, $(-300\ \mu m, 300\ \mu m, 300\ \mu m)$, $(-300\ \mu m, -300\ \mu m, 300\ \mu m)$, and $(300\ \mu m, -300\ \mu m, 300\ \mu m)$ with an equal weight factor of unit.

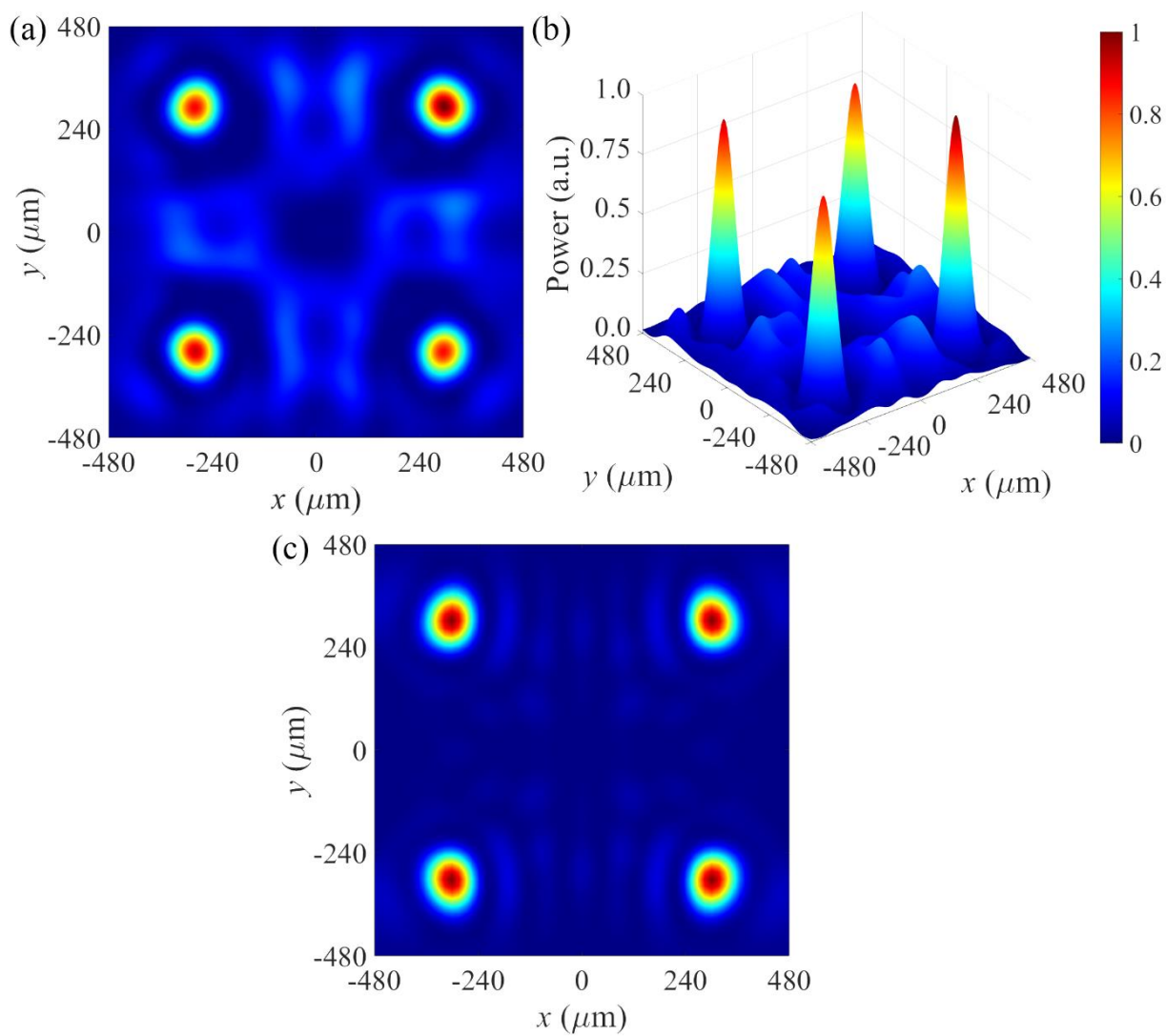

*Supplementary Figure S4* – Simulated (a) Two-dimensional and (b) three-dimensional and (c) theoretical two-dimensional power distribution on the focal plane for the fourth example.

iii